# Spin-filtering in superconducting junction with the manganite interlayer

G.A. Ovsyannikov, K.Y. Constantinian, V.V. Demidov, Yu.V. Kislinskii, A.V. Shadrin and A.M. Petrzhik

*Kotel'nikov Institute of Radio Engineering and Electronics RAS, Moscow, Russia*

Abstract

We report on the electronic transport and the impact of spin-filtering in mesa-structures made of epitaxial thin films of cuprate superconductor $YBa_2Cu_3O_x$(YBCO) and the manganite $LaMnO_3$ (LMO) interlayer with the Au/Nb counterelectrode. Ferromagnetic resonance measurements of heterostructure Au/LMO/YBCO shows ferromagnetic state at temperatures below 150 K as in the case of reference LMO film grown on the neodymium gallate substrate. The heights of the tunneling barrier evaluated from resistive characteristics of mesa-structures at different thickness of interlayer showed an exponential decrease from 30 mV down to 5 mV with the increase of manganite interlayer thickness. Temperature dependence of the conductivity of mesa-structures could be described taking into account the d-wave superconductivity in YBCO and a spin filtering of the electron transport. Spin filtering is supported also by measurements of magneto-resistance and the high sensitivity of mesa-structure conductivity to weak magnetic fields.

1. Introduction

Interaction between superconductivity and magnetism is one of the important areas of research in solid state physics. The problem has been studied for magnetic superconductors as well for hybrid structures containing superconducting (S) and ferromagnetic materials (F) [1-3]. The superconducting junctions with a interlayer of a ferromagnetic insulator (S/FI/S) have intriguing potential for practical applications in spintronics for realization of π-junction [4] and unusual phase dynamics of superconducting current [4-6]. The tunnel current flowing through the ferromagnetic layer between two non-magnetic electrodes becomes spin-polarized due to exchange splitting of the ferromagnetic insulator band with the states with the spin up and spin down. Exchange splitting is proportional to exchange energy $E_{ex}$ of ferromagnetic interlayer. Since the tunneling probability increases exponentially with the reduction of barrier height a S/FI/S junction can effectively produce almost a fully spin polarized current [7-8]. The magnetic barrier affects also the phase difference of electrons with spins up and down [9]. The both spin-filtering and spin-mixing have significant impact on electron transport characteristics in the junction with magnetic interlayer. The mechanical stress of the films and the processes of electronic adjustment at the interfaces substantially affect the parameters of heterostructures with magnetic barrier [10]. Spin-filtering effects have been studied extensively in ferromagnetic insulating europium chalcogenides [11-13]. However, low Curie temperatures ($T_{CU}$) and poor chemical compatibility with potential electrodes limit their application in spin filtering. Results of the investigation of hybrid structures containing superconducting cuprate $YBa_2Cu_3O_x$ (YBCO) and manganite $LaMnO_3$ (LMO) which have good chemical and crystal structure compatibility are presented. An interface of epitaxial layers manganite/cuprate superconductor $La_{0.7}Ca_{0.3}MnO_3/YBa_2Cu_3O_x$ (LCMO/YBCO) could be produced almost free of defects [14-17] and only manganese ions migrated within distance of 1 nm [18]. These properties of manganite-cuprate interfaces were used [19] for fabrication of our heterostructures. Niobium was chosen as a top superconducting electrode with a buffering Au film. All-oxide heterostructures YBCO/LMO/YBCO were studied in [20] where electron transport at high voltages $V \sim 1$ V was investigated. Although in the [20] it was pointed on possibility of spin filtering in heterostructures with LMO interlayer, an experimental evidence for it has not been shown.

The potential of application of manganite ferromagnetic insulator for spin-filtering has been demonstrated in [21], where an ultrathin film of $Sm_{0.75}Sr_{0.25}MnO_3$ was integrated as a interlayer in epitaxial oxide tunnel junction showing spin polarization up to 75% at 5 K.

2. Experimental samples

The heterostructure LMO/YBCO was fabricated by laser ablation at temperature 700-800°C and pressure 0.3 mbar on a substrate (110)$NdGaO_3$ (NGO). Thin ($d_M = 5-20$ nm) LMO film was epitaxially grown after deposition of YBCO film with thickness 100-150 nm. After cooling to room temperature a thin (20-30 nm) layer of Au was deposited in situ. A subsequent layer of niobium (Nb) was deposited by magnetron sputtering in other chamber. The buffering Au film is used to reduce oxygen diffusion at the LMO/Nb interface. The proximity effect between Nb and Au films provided bilayer critical temperature $T_C' = 8.5-9$ K that is close to the critical temperature of bulk Nb (9.2 K). The critical temperature of the YBCO film was $T_C = 88-89$ K. The stoichiometric LMO manganite at low temperatures $T$ has properties of an insulator and should be antiferromagnetic [22-24]. The layout of micron-sized mesa-structures made from Nb-Au/LMO/YBCO heterostructers were formed by photolithography, plasma-chemical and ion-beam etching [25, 26]. Mesa-structures had square shape with linear dimensions $L$=10-50 μm.

3. Experimental results and discussion

3.1 Thin films of LMO

Fig. 1 shows the temperature dependence of the resistivity $\rho$ of autonomous LMO epitaxial films deposited on a NGO substrate. Note that the resistance of LMO film is significantly higher than the resistance of doped manganite ferromagnetic films, particularly,

La$_{0.7}$Sr$_{0.3}$MnO$_3$. The increase in resistance with decreasing the temperature indicates non-metallic conductivity. A detailed analysis of the temperature dependence shows that there are two parts of $\rho(T)$ dependence that are typical for the Mott insulator, which are described by $\rho \propto \exp(T_0/T)^{1/4}$ with different characteristic temperature $T_0 = 34\ 10^6$ K at $T > T_{CU}$ and $T_0 = 4\ 10^6$ K at $T < T_{CU}$, where $T_{CU}$ is phase transition temperature (later will be shown that it is Curie temperature). The difference for $T_0$ below and above $T_{CU}$ could be described by polaronic hoping model at high temperature [21, 27] for which the temperature dependence of resistance looks like $\rho = AT\exp(T_0^*/T)$, where $T_0^*$ is activation temperature and $A$ is depended on carrier concentration and hoping length.

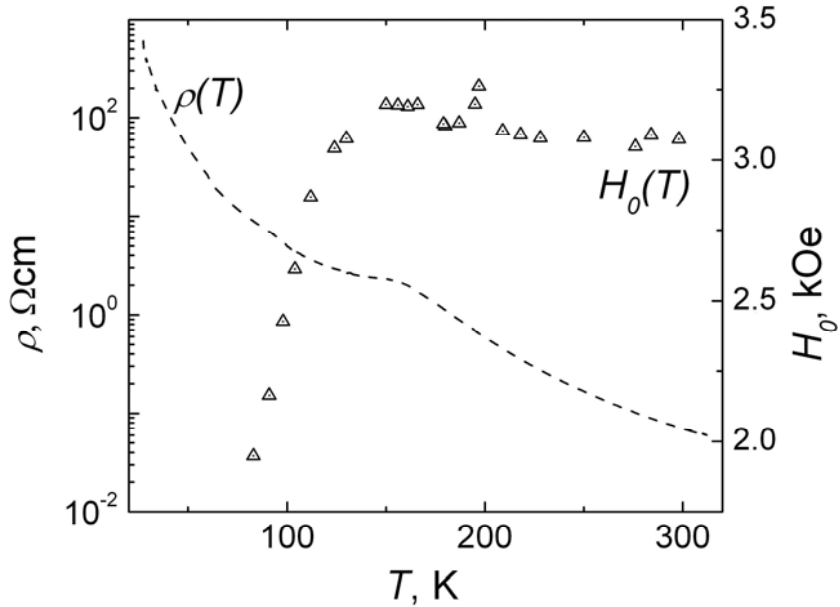

Fig.1. The temperature dependence of the autonomous LMO film resistivity $\rho(T)$ (the dashed line). The temperature dependence of a resonance magnetic field $H_0$ is shown by triangles

Magnetic properties of LMO films and heterostructures were determined from the spectra of magnetic/ferromagnetic resonance (FMR) at a frequency of 10 GHz [28-30]. The weak growth of resonant field $H_0$, (the paramagnetic phase) at high temperatures (just below the room temperature) is replaced by a sharp decrease $H_0(T)$ in the vicinity of the Curie temperature $T_{CU} \approx 150$ K (Fig. 1). The obtained $T_{CU}$ values differs from the results of the magnetic susceptibility measurements of single-crystal samples and depended significantly on the crystalline quality of the films, the strain and the of oxygen contents [29, 30]. The $T_{CU} \approx 150$ K obtained from FMR data is close to the temperature where a changing of curvature of $\rho(T)$ takes place (see Fig. 1 dashed line). The magnetization could be estimated from $H_0(T)$. For T < 100 K magnetization $M_0 \sim 2$ $\mu_B$/Mn was obtained.

It is known that the reason of ferromagnetism in manganites is caused by double exchange between $Mn^{3+}$ and $Mn^{4+}$ [24]. For manganites with low doping an antiferromagnetic phase is expected due to superexchange interaction. An important role in the interaction of $Mn^{3+}$ ions plays Jahn-Teller distortion [29, 31, 32]. However, the ferromagnetic phase is observed even in a stoichiometric $LaMnO_3$ films with a very small change in the composition of the oxygen. The experimentally determined value of the exchange interaction energy obtained from neutron measurements of LMO crystals ($E_{ex} \sim 1$ meV) indicate low values amplitude and anisotropy on *ab* and directions LMO crystal because of presence both a double exchange and superexchange interactions [33, 34]. The magnetic polarization of the films LMO, estimated by comparing the intensities of the FMR signals is close to the magnetic polarization of $La_{0.7}Sr_{0.3}MnO_3$s film which reaches 100% at low temperatures [35, 36].

3.2. The conductivity of the mesa-structure with a LMO interlayer

Inset in Fig. 2a presents cross-section of the mesa-structure, dc current supply (I) and voltage measurements (V). More than ten mesa structures were fabricated and measured. Typical mesa structures parameters are shown in Table 1.

Table 1. The thickness and resistance area mesa structures at 4.2 K

| N | $d_M$, nm | $A$, $\mu m^2$ | $R$, $\Omega$ | $RA$, $\mu\Omega$ $cm^2$ |
|---|---|---|---|---|
| 1 | 6 | 360 | 66 | 238 |
| 2 | 6 | 140 | 164 | 230 |
| 3 | 8.5 | 680 | 44 | 300 |
| 4 | 8.5 | 360 | 86 | 310 |
| 5 | 17 | 1600 | 125 | 2300 |
| 6 | 17 | 400 | 258 | 1032 |

$d_M$ is thickness of LMO interlayer, $A$ and $R$ are area and resistance of the mesa structure correspondingly.

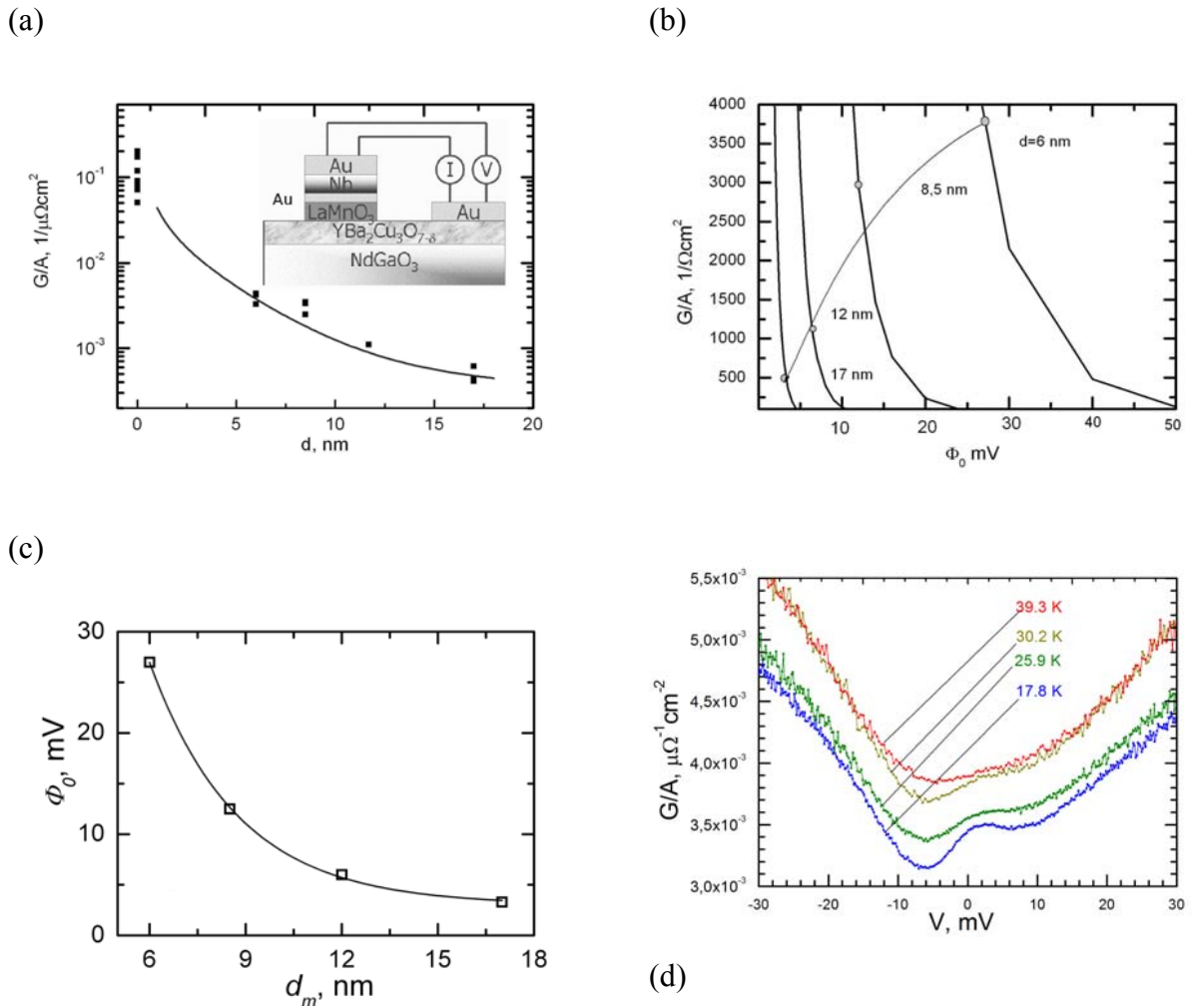

Figure 2. a) LMO interlayer thickness ($d_M$) dependence of mesa structure characteristic conductivity $G/A$, measured at $V=0$, $T=4.2K$. The solid line is the dependence calculated by (1), taking into account the changes in the barrier height from thickness dependence (Fig.2c). The squares are experimental data. The inset shows cross-section of mesa-structure and current supply (I) and voltage measurements (V) circuits, (b) the determination of the mean barrier height $\Phi_0$ from characteristic conductivity, (c) the dependence of $\Phi_0$ from $d_M$, (d) voltage dependence of the conductivity at $T=17.8$ K, 25.9 K, 30.2 K and 39.3 K.

Fig. 2a shows LMO layer thickness dependence of the characteristic conductivity $G/A$ at $V=0$, $T=4.2K$ ($A$ is area of the mesa-structure). With decreasing thickness $d_M$ the contribution from the tunnel barrier becomes decisive and experimental G/A values are significantly higher than those, $G_0$, calculated from the conductivity of LMO interlayer. It should be noted in the structures without the LMO interlayer (data for $d_M = 0$) the barrier height between Au and YBCO is much higher than for Au/LMO/YBCO heterostructure case. Using the model for tunneling current [37, 38] in the tunnel contacts of two metals we obtain for the conductivity $G_0/A$ of mesa-structure at

low voltages and rectangular barrier:

$$G_0 / A = (e/h)^2 (2me\Phi_0 / d_M^2)^{1/2} \exp(-2d_M \sqrt{2me\Phi_0} / h) \qquad (1)$$

$\Phi_0$ is the average height of the barrier, e and h are the electron charge and Planck constant correspondingly, $d_M$ is interlayer thickness. Curve $G_0/A$ vs. $d_M$ in Fig 2a was obtained using thickness dependent $\Phi_0(d_M)$. Fig.2b shows a family of $G_0/A$ vs. $\Phi_0$ calculated by (1) for fixed experimental $d_M$, points are the experimental conductance G/A values averaged over a few samples with the corresponding $d_M$. It's seen that an exponential approximation curve crosses experimental points except the case of $d_M$=8.5 nm. $\Phi_0(d_M)$ dependence is shown in Fig. 2c. Obtained by fitting the experimental data the barrier height does not exceed 30 mV, which is significantly smaller than the barrier heights for $Al_2O_3$, MgO and other traditional insulators

a)

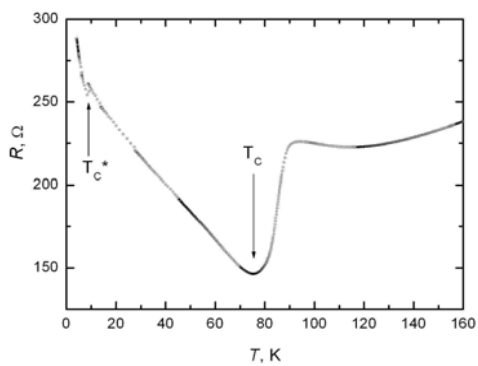

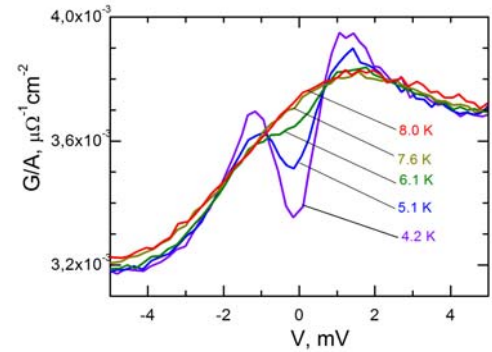

b)

Fig 3. Temperature dependence of the resistance for mesa-structure N6. The superconducting transition temperature of YBCO $T_C$ = 78 K and the bilayer Nb-Au $T_C^*$ = 8.7 K. (b) voltage dependence of G/A(V) for T=4.2, 5.1, 6.1, 7.6, 8.0 K.

Temperature dependence of resistance R of mesa structure N6 is shown in Fig. 3a. At high temperatures $T$> 150K $R(T)$ is determined mainly by resistance of YBCO electrode. Resistance drops near $T_C$ caused by superconductivity transition of YBCO film, and then near $T_c^*$ by superconductivity of Nb/Au bilayer. For mesa-structures with thicknesses $d_M$ = 5-20 nm superconducting current was absent, while in the case of $d_M$ = 0 clear superconductivity was observed [26]. Fig.3b shows evolution of differential conductance when temperature was changed from 4.2 K to 8 K at $T<T_C^*$. At $T <T_C^*$ mesa-structures could be considered as $S/FI/S_d$ structure in which the LMO is a magnetic barrier (FI), bilayer Nb-Au is a superconductor with the s-symmetry of order parameter and the opposite electrode YBCO is a superconductor ($S_d$), in which the dominant contribution comes from d-wave symmetry of order parameter.

It is seen that $G(V)$ strictly depend on temperature at low voltages. An asymmetry over bias voltage is an other feature which is observed in wider voltage range.

At higher temperatures $T_C^* < T < T_C$ mesa-structure can be considered as a N/FI/S$_d$, because Nb-Au bilayer is in the normal state. In the case of a superconductor with the s-wave symmetry of the order parameter at temperatures much lower than the critical, only a small number of quasiparticles are above the gap. At $T = 0$ there are almost no excited quasiparticles above the gap ($\Delta/e$) and the tunneling current for NIS structure is close to zero. As a result, at temperatures below the critical one, we have an exponential dependence of the heterostructure conductivity on $G \propto \exp(-\Delta/T)$. For a superconductor with a d-wave symmetry as shown by calculations [39] the quasiparticle current in the heterostructure N/I/S$_d$ is not small at voltages $V << \Delta_d/e$. At low voltages for N/I/S$_d$ structures a theoretical conductivity $G$, normalized on conductivity at $T=T_C$, $g_S$, is given by the following equation

$$g_S = (k_B T/e\Delta_d)\ln(e\Delta_d/kT) \qquad (2)$$

where $k_B$ is Boltzmann's constant. Fig. 4 shows the normalized temperature dependence of the experimental conductivity g of the two mesa structures N3 and N5 with thicknesses of layers 17 and 8 nm. The temperature dependence (2) predicts reduction of heterostructure conductivity with decreasing $T$. In Fig.4 it is shown by curve 2, calculated taking into account the contributions from S$_d$ electrodes, and taking $e\Delta_d/k_B T_C=5$. There is apparent deviation of experimental data from theory [39] which predicts much rapid falling at low temperatures.

In the presence of ferromagnetism in the interlayer the height of the barrier is different for electrons with spin up and down on the amount of $E_{ex}$: $\Phi_{\uparrow,\downarrow} = \Phi_0 \pm E_{ex}/2$. Because of the exponential dependence of the tunneling current from barrier height [37, 40], there is a strong difference between currents with different directions of the spins [38].

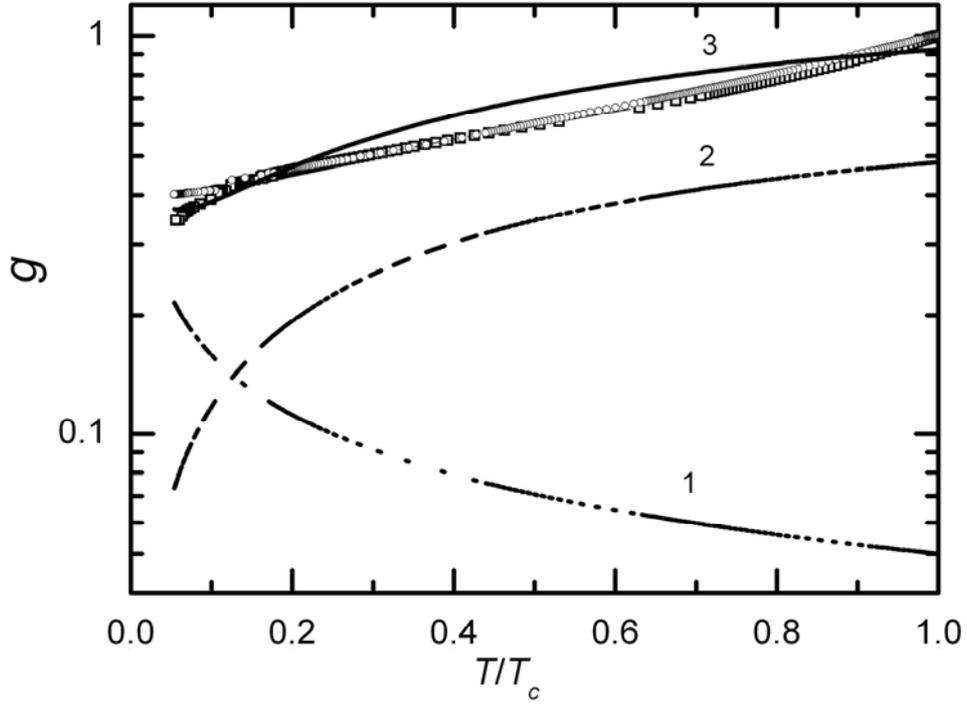

Figure 4. Temperature dependence of normalized conductivity for the mesa structures N3 (filled squares) and N5 (circles) respectively. Curve (1) shows the relationship (3), the dashed line dependence (2), and a solid line (3) shows the sum with coefficients that ensure the best agreement with experiment.

Calculations of the spin polarization effect on the conductivity of N/FI/S structures with ferromagnetic tunnel layer (FI) have shown that the conductivity of the structure is determined by effectivity of interlayer spin filtering $P_b = (T_\uparrow^2 - T_\downarrow^2)/(T_\uparrow^2 + T_\downarrow^2)$, where $T_{\uparrow(\downarrow)}$ is the probability of transport for the tunneling current with spin up (down) [42], The normalized values of the heterostructure conductivity at zero voltage is given by the equation:

$$g_M(T) = 0.21 \chi_T \sqrt{1 - P_b^2} \sqrt{\frac{hD}{2kT}} \qquad (3)$$

where $D = v_f l/3$ is diffusion coefficient, $v_f$ is Fermi velocity, $l$ is mean free path, $\chi_T = \rho_S/RA$, $\rho_S$ is the normal resistivity of the superconducting electrode. Since in our mesa-strictures $P_b < 1$ can only grow slightly with lowering $T$ from the Curie temperature of the ferromagnetic layer LMO, the contribution to mesa-structure conductivity due to the spin filter should increase at $T < T_{CU}$ inversely proportional to square root of $T$ (3). As a result, it can be assumed that the combined effect of d-wave superconductivity reduces conductivity mesa-structure (2), while temperature dependence of spin filtering, increases conductivity (3) with temperature lowering, giving an explanation of the experimental dependence of $g(T)$.

3.3. Magnetic properties

To obtain magnetic parameters of heterostructures the magnetic resonance spectra studies were carried out at $f = 9.56$ GHz. All spectra were recorded in a parallel orientation when the external magnetic field was directed parallel to the substrate plane. Magnetic resonance signals were observed when the magnetic ordering can only occur in the LMO film.

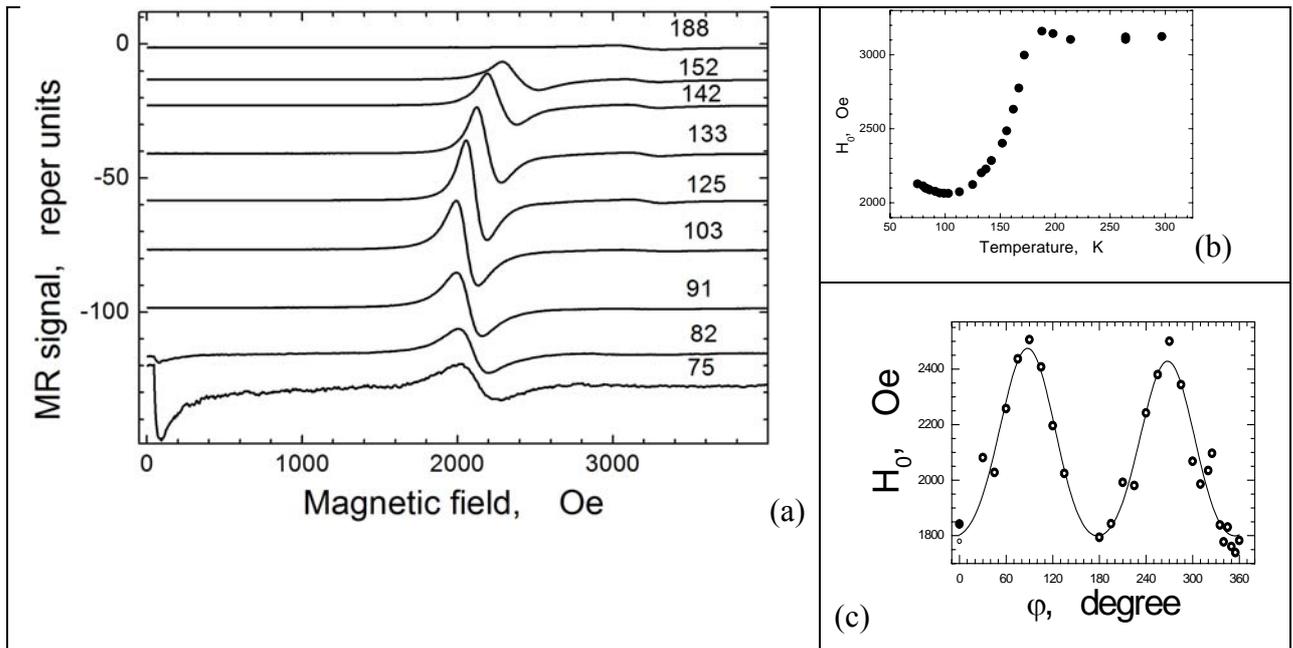

Fig.5. (a) FMR spectra for temperatures range 188-75K. (b) temperature dependence of ferromagnetic resonance field for heterostructure Au/LMO/YBCO with interlayer thickness of LMO film $d_M = 17$ nm (c) angular dependence of resonance field $H_0$ at $T=115$ K.

Fig. 5a shows magnetic resonance spectra of heterostructures Au/LMO/YBCO/NGO for temperatures range 75-188 K. It is seen that, magnetic resonance line is shifted into the low field with decreasing temperature. The signal intensity is increased significantly in the range of 160-300 K. In addition, non-resonant absorption signal at low magnetic fields (see $T = 75$ K) which is an indicator of the superconductivity in YBCO film appear. Fig. 5b shows the temperature dependence of the resonance field which is characteristic for the behavior of the spectrum of the film structures with increasing magnetization due to ferromagnetic ordering at $T < T_{CU}$. The Curie temperature is accepted as the value corresponding to the maximum of the derivative for the dependence shown in Fig. 5b. Fig. 5c shows the angular dependence of the resonant field values at $T = 115$ K. The values of uniaxial $H_u$ and cubic $H_c$ magnetic anisotropy, their direction and the equilibrium magnetization $M_0$ of ferromagnetic LMO were calculated from the best fitting of angular dependence, $H_0(\varphi)$ [28, 29]. The values $H_u$ =375 Oe, $H_c$=45 Oe, and $M_0$=1.75$\mu_B$/Mn, and the axis of easy magnetization deployed on 45° relative to each other were obtained. These parameters are typical of manganite films grown on NGO substrates [28, 29].

For further characterization of spin polarization of the barrier we studied magnetic field dependences of the mesa-structures conductivity. Fig. 6a shows the voltage dependence of mesa-structure differential resistance $R_d(V)$ at $T$ = 4.2K for a few fixed external magnetic fields. It is seen an asymmetry of $R_d(V)$ in the voltage range $|V| \leq$ 5mV. According to [37, 40], the asymmetry relative to $V$=0 could be caused by a scalene trapezoid shape of the tunnel barrier due to different work functions for YBCO and Au. Since the mean barrier height $\Phi_0$ is less than 30 mV (see Fig. 2a) the asymmetry in Rd (V) is observed at voltages less than $\Phi_0$. At voltages $V> \Phi_0$ the observed increase of shot noise indicate on tunnel conductivity of the mesa structure (see. inset in Fig. 6a). When relatively weak external magnetic field was applied the $R_d(V)$ changes in the voltage range $|V| \leq \Delta_{Nb}/e$ , which can not be associated with the Zeeman splitting of the energy states of electrode [41]. This is a direct indication of spin polarization impact on the tunneling current [42]. External magnetic field directly affects the polarization, and therefore the components of tunneling conductance depended on imbalance of $\Phi_\uparrow$ and $\Phi_\downarrow$ [41]. When the voltage is greater than the height of the tunnel barrier, there is no spin filtering and ceases to operate the associated spin-dependent tunneling conductance.

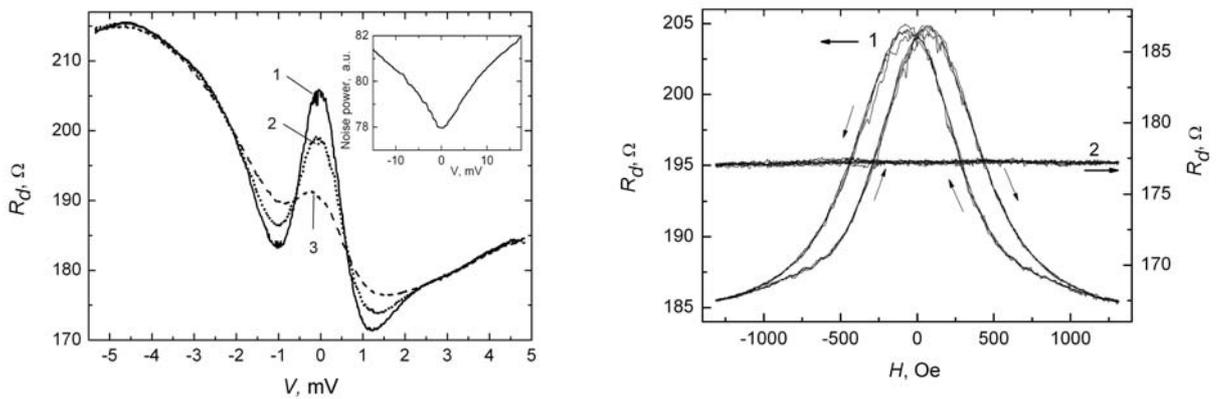

Fig. 6. (a) The voltage dependence of differential resistance for mesa-structure N2 $T$ = 4.2K for the external magnetic fields $H$ = 1- 0, 2 - 263 Oe, 3- 526 Oe. The magnetic field is parallel to the substrate plane. The inset shows shot noise dependence. (b) Magnetoresistance of mesa-structure N 2 at $T$ = 4.2 K. The arrows indicate the direction of the magnetic field changing. The curves (1) obtained in at $V$ = 0 for central peak in Fig.6a, the curves (2) was obtained for bias voltage $V$ = 10 mV, where spin-filtering no longer observed.

Fig. 6b shows the magnetoresistance of the mesa-structure N2 under external magnetic field directed parallel to the plane of the substrate. The hysteresis, that is characteristic for ferromagnetic systems by changing the direction of the magnetic field, was clearly observed. The magnitude of the magnetoresistance of a few tens of percent is typical for structures with manganites [28]. By increasing the DC bias voltage to 10 mV the differential resistance of mesa-structure does not depend on the magnetic field (curve 2 in Fig. 6b). So, we observed magnetoresistance of mesa-structure, not of manganite. Otherwise the magnetoresistance should not depend on voltage biasing.

4. Conclusion

Thus, the study of the ferromagnetic resonance in epitaxial hybrid superconducting heretostructures of cuprate superconductors and gold layer with a interlayer of manganite shown experimentally that the heterostructure goes into a ferromagnetic state at a temperature of 150-160 K as in the case of LMO autonomous film. From the experimental dependence of the characteristic resistance of the mesa structures on the thickness of interlayers the barrier height is obtained. To explain the experimental temperature dependence of the conductivity of the mesa structure in the temperature range between the critical temperatures of superconductors d-wave superconductivity for one of the electrodes and the spin filtering should be taken into account. Spin filtering in heterostructure was confirmed by the high sensitivity to external magnetic field and hysteresis of the magnetoresistance for mesa structure in the bias voltage not exceeding gap of s-wave superconductor.

We thank A.S. Vasenko, A.F. Volkov, A.S. Grishin for help and useful discussions. The work was supported in part by grants RFBR 16-37-60069, 16-19-14022 and scientific school NSH-8168.2016.2 and with the assistance of the Swedish national center equipment for micro and nano technology (Myfab).